\documentclass[conference]{IEEEtran}

\usepackage[colorlinks,urlcolor=blue,linkcolor=blue,citecolor=blue]{hyperref}

\usepackage{makecell, subcaption}
\usepackage{amsmath,amssymb,amsfonts}

\usepackage{ragged2e}
\usepackage{algorithm}
\usepackage{algcompatible}

\usepackage{graphicx}
\usepackage{textcomp}
\usepackage{rotating}
\usepackage{balance}
\usepackage{multirow}
\usepackage{gensymb}
\usepackage{array,booktabs,longtable,tabularx,tabulary}
\usepackage{lipsum, balance}

\usepackage{hyphenat}

\newcommand*{\email}[1]{\normalsize\texttt{\href{mailto:#1}{#1}}\par}

\usepackage{threeparttable, booktabs, url, eqparbox, xspace, setspace, tabularx, color, soul, wrapfig}
\makeatletter
\DeclareRobustCommand\onedot{\futurelet\@let@token\@onedot}
\def\@onedot{\ifx\@let@token.\else.\null\fi\xspace}

\def\etal{\emph{et al}\onedot}

\begin{document}

\title{Detection of Physiological Data Tampering Attacks with Quantum Machine Learning}



\author{
    \IEEEauthorblockN{%
         Md. Saif Hassan Onim and
         Himanshu Thapliyal 
                    }
    \IEEEauthorblockA{%
        Department of Electrical Engineering and Computer Science\\
        University of Tennessee, Knoxville, TN-37996, United States}
    
    \email{monim@vols.utk.edu},
    \email{hthapliyal@utk.edu}
                
    }
\maketitle

\begin{abstract}
The widespread use of cloud-based medical devices and wearable sensors has made physiological data susceptible to tampering. These attacks can compromise the reliability of healthcare systems which can be critical and life-threatening. Detection of such data tampering is of immediate need. Machine learning has been used to detect anomalies in datasets but the performance of Quantum Machine Learning (QML) is still yet to be evaluated for physiological sensor data. Thus, our study compares the effectiveness of QML for detecting physiological data tampering, focusing on two types of white-box attacks: data poisoning and adversarial perturbation. The results show that QML models are better at identifying label-flipping attacks, achieving accuracy rates of $75\%-95\%$ depending on the data and attack severity. This superior performance is due to the ability of quantum algorithms to handle complex and high-dimensional data. However, both QML and classical models struggle to detect more sophisticated adversarial perturbation attacks, which subtly alter data without changing its statistical properties. Although QML performed poorly against this attack with around $45\%-65\%$ accuracy, it still outperformed classical algorithms in some cases.
 
\end{abstract}

\begin{IEEEkeywords}
data tampering, SVM, adversarial attack, machine learning, threat model
\end{IEEEkeywords}

\section{Introduction}
\label{intro}
In contemporary healthcare, wearable and implanted devices are becoming widespread for monitoring physiological parameters. These devices frequently connect to cloud-based platforms enabling faster decision making. However, the increased reliance on networked systems has prompted significant concerns regarding data integrity, particularly in the context of cyber risks such as tampering attempts. Data tampering, in which an adversary manipulates physiological data, can lead to inaccurate diagnoses and potentially hazardous medical decisions. If these systems get compromised, it could lead to disastrous incidents involving health and safety~\cite{malik2024}. 
Unattended devices are vulnerable to physical harm by hackers, leading to data alteration invasions. In most cases, hackers can modify software, install malware, swap out the device for a malicious clone, control sensors, change hardware, and more~\cite{li2010}. A compromised sensor could transmit erroneous information to the gateway node, resulting in inaccurate diagnoses and treatments that could endanger the patient~\cite{dimitriou2008}. Machine learning has long been used to solve such tampering detection problems~\cite{pathak2021, pacheo2018, montalvo2023}. However, with the increasing complexities of datasets and ML models, classical algorithms are starting to fall short. Again the adversary or hackers benefit from easily available resources to further complicate the detection process~\cite{sun2022, yerlikaya2022}.

Now with emerging technologies like Quantum Machine Learning (QML), the problems can be re-evaluated from different perspectives. It uses quantum computing concepts to improve existing machine learning approaches and can detect subtle patterns in tampered data that may bypass traditional methods~\cite{qml, Havlicek2019}. Due to their enhanced capacity for pattern recognition, quantum algorithms like quantum support vector machines have the potential to greatly improve the detection of anomalies in a dataset. Cultice~\etal~\cite{cultice2024} showed the advantages of QML in anomaly detection in cyber-physical systems. However, the performance of QML in physiological data tampering detection is still an under-explored domain with several open research questions.

To address these questions, this paper explores the application of hybrid QML in detecting tampering attacks on physiological data. We also compare the effectiveness of QML-based approaches with classical machine learning techniques. Our contributions are as follows:

\begin{itemize}
    \item We experimented with three types of physiological sensor datasets with one, two, and three class labels respectively.

    \item We explored the performance of a threat model that performs two types of white box attacks on physiological sensor data.
    
    \item We utilize a quantum hybrid framework for the detection model that uses kernel-based SVM with a quantum kernel circuit.

    \item We report a detection accuracy of around 75\% to 95\% for data poisoning attacks for our hybrid QML framework. We also show the comparison of performance in detection between classical and quantum machine learning models.

\end{itemize}

Based on the contemporary literature, this is the first work in this domain that utilizes quantum machine learning to detect physiological data tampering. We have mentioned our motivation and contributions in Section~\ref{intro}. The rest of this paper has the following structure: Section~\ref{threat} discusses briefly the threat model and the attack. Section~\ref{method} describes the framework of our Hybrid QML detection model and its training. Section~\ref{result} contains information about our experimental result. Section~\ref{Conclusion} wraps up the report by suggesting potential future directions for this research.

\section{Threat Model}
\label{threat}
A threat model defines the risk conditions and their associated danger distributions~\cite{xu2012}. Here, we define risk conditions based on the likelihood and severity of adversarial tampering in cloud-connected physiological monitoring systems. The primary risk conditions include unauthorized data access, model poisoning via malicious training data, and hardware-level vulnerabilities. Their danger distributions describe how different adversarial attacks impact system integrity. For example, minor statistical deviations in physiological data can go undetected and cause misclassification without triggering system-wide alerts. Our threat model includes two forms of white box attacks: \textit{Data Poisoning} and \textit{Adversarial Perturbation}. According to the model, the adversary acquires unauthorized access to the data stream while being transferred to the cloud. This allows them to initiate Adversarial Perturbation and execute a Label Flipping attack. Thus, the model trained on adversarial examples will degrade in performance. We will now briefly describe the attacks.


\subsection{Data Poisoning}
Data poisoning attacks occur when an adversary intentionally manipulates the training data of a machine learning model, deteriorating its performance. Label flipping, a typical method of data poisoning, involves an attacker changing the labels of specific training samples while leaving their features unaltered. In targeted poisoning, the attacker attempts to trick the model into creating specific mistakes, usually against particular instances or classes.

\subsection{Adversarial Perturbation}
An adversarial perturbation attack takes advantage of the vulnerability of machine learning models by introducing small, carefully constructed perturbations to input data, causing the model to make inaccurate predictions~\cite{moosavi2017universal}. The attacker aims to generate an adversarial example $x = x +\delta$, where $\delta$ is a small perturbation added to the original input $x$, such that: $ f_{\theta}(x') \neq f_{\theta}$.

\section{Proposed Tampering Detection Method}
\label{method}

\begin{figure}
    \centering
    \includegraphics[width=\columnwidth]{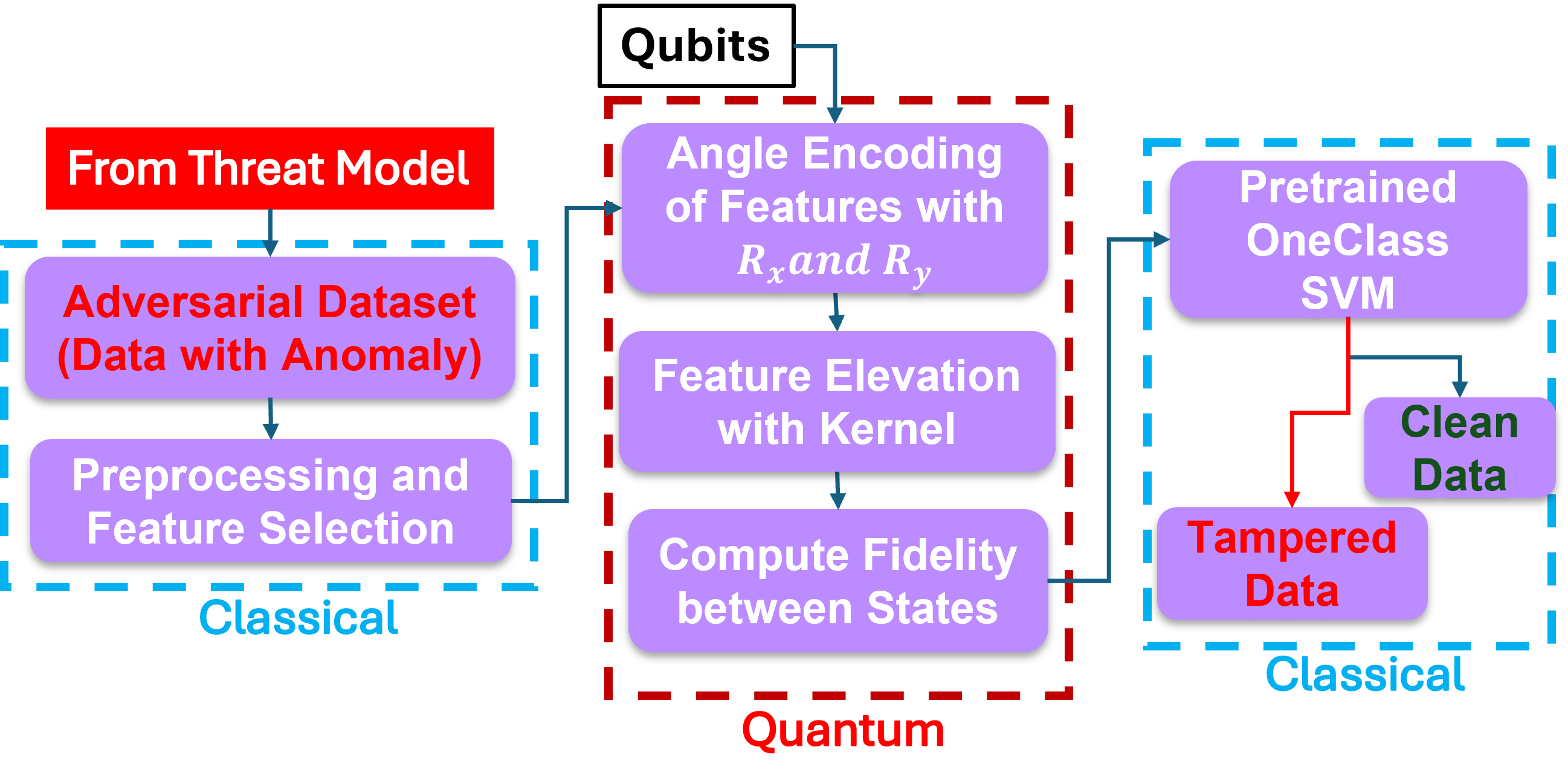}
    \caption{Workflow of our Proposed Hybrid QML model for Tampering Detection}
    \label{fig:workflow}
\end{figure}

Our tampering detection approach uses Quantum Hybrid One-Class SVM to address data tempering as an anomaly detection problem. The detection model proceeds in three steps. In the classical domain, it first performs both preprocessing and feature reduction on the data sent from the threat model. Next, a quantum machine calculates quantum fidelity from the kernel. Lastly, the classical domain is used to optimize the One-Class SVM. Figure~\ref{fig:workflow} illustrates our proposed method's overall workflow. This method's SVM-based design utilizes quantum fidelity to map the feature space and identify similarities in data. We will now describe these steps in detail.

\subsection{Dataset Preprocessing and Feature Extraction}
\label{dataset}
We selected three types of datasets having one, two, and three labels for our experiment. Table~\ref{tab:dataset} shows a summary of their descriptions.
\begin{table}[htbp]
  \centering
  \caption{Description of Dataset}
  \renewcommand{\arraystretch}{1.5}
    \setlength{\tabcolsep}{6pt}
    \resizebox{.98\columnwidth}{!}{
    \begin{tabular}{ccccc}
    \toprule
    \textbf{Dataset} & \textbf{Signals} & \textbf{Class} & \makecell[c]{\textbf{Participants}\\\textbf{Age}} & \makecell[c]{\textbf{Extracted}\\\textbf{Features}} \\
    \midrule
    \makecell[c]{RESTINGECG\\~\cite{restingecg}} & ECG & One & \makecell[c]{22\\(18-26)} & 12\\
    \midrule
    \makecell[c]{EPHNOGRAM\\~\cite{ephnogram, PhysioBank}} & ECG, PCG & Two & \makecell[c]{24\\(23-29)} & 24\\
    \midrule
    \makecell[c]{Stress\\~\cite{nath2022, onim2024utilizing, onim2024predict}} & \makecell[c]{EDA, BVP,\\ST, IBI} & Three & \makecell[c]{40\\(60-80)} & 60\\
    
    \bottomrule
  \end{tabular}
  }
  \label{tab:dataset}
\end{table}

RESTINGECG~\cite{restingecg} has one class that represents the ECG signal stream in resting conditions. They recorded continuous EEG signals across 72 channels with electrodes. All channels were amplified in 24-bit DC mode, which was initially sampled at 2,048 Hz (400 Hz bandwidth) and subsequently downsampled online to 256 Hz. EEG signals were captured using a common mode sense (CMS) electrode. Half-cell potentials at the electrode/gel/skin interface were kept within 40 mV, following established standards.

The EPHNOGRAM~\cite{ephnogram, PhysioBank} project has two class labels, namely \textit{normal} and \textit{anomaly}. It aimed to develop low-cost, low-power devices capable of simultaneously recording electrocardiogram (ECG) and phonocardiogram (PCG) data and additional channels to capture environmental audio noise. The current dataset, recorded using version 2.1 of the hardware, was collected from 24 healthy persons. 

The Stress dataset~\cite{onim2024utilizing, onim2024predict} has three labels: \textit{not-stressed, low-stress} and \textit{high-stress}. Here, 40 older adults aged 60 to 80 were chosen as the target demographic. To provide an unbiased dataset, each subject completed a thorough screening process before participating in the experimental methodology to rule out any pre-existing medical issues. Physiological data were gathered using the Empatica E4 wristband, which includes sensors for skin temperature (ST), photoplethysmography (PPG), and electrodermal activity (EDA) while they participate in TSST (Trier Social Stress Test). Their cortisol samples were collected from their saliva from a time interval to be used as a labeling scheme. 

12 statistical features are retrieved from each time domain signal in the EPHNOGRAM and RESTINGECG datasets making it a total of 24 Features for EPHNOGRAM and 12 for RESTINGECG. From the Stress dataset, 60 features were extracted.

\subsection{Sampling and Feature Reduction}
Due to the limitation of resources in terms of memory and runtime, we downsampled the dataset with a moving average window of 60 samples. It helps reduce the noise and also captures the time series patterns. The average samples are calculated using $avg_{t}=\frac{1}{w}\times\sum_{i=t-w+1}^{t}x_{i}$. Here, $w$ is the window length and $x_i$ is the sample at time $i$. Next, the categorical samples are substituted with numerical histogram values that are better suited to the ML algorithms. The dataset is then standardized with Z-score normalization by $x_{std}=\frac{x_{o}-\mu}{\sigma}$ where, $x_{0}$ is the original sample, $\mu$ is the sample mean and $\sigma$ is the standard deviation. This moves the mean of the distribution to zero and the standard deviation to one. Furthermore, standardization tends to accelerate convergence for optimization techniques such as gradient descent. Again, the computational complexity and runtime depend mainly on the number of qubits. With each number of additional qubits, the complexity increases quadratically. For our experiment, we decided to use 6 qubits that can hold 12 features. Thus, we used PCA (Principal Component Analysis)~\cite{pca} as a feature reduction technique.

\subsection{Encoding and Feature Map}
\begin{figure}[htbp]
    \centering
    \includegraphics[width=\columnwidth]{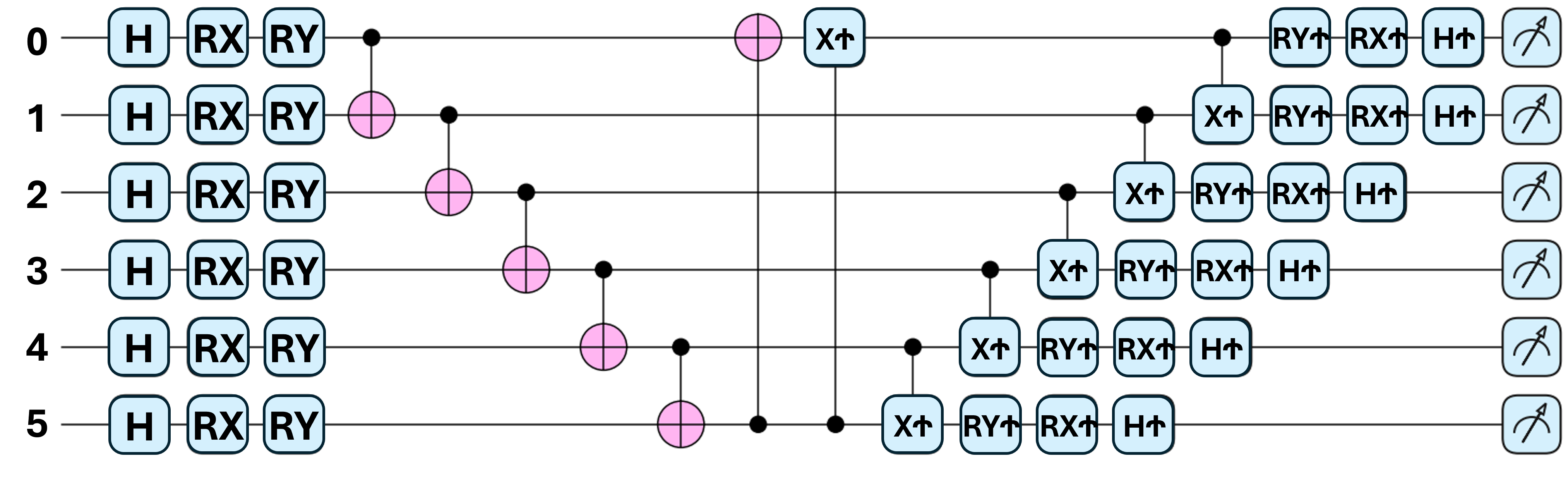}
    \caption{Kernel Computation Circuit. ($0-5$) are the qubits, $H$ is the Hadamard gate, $R_x$ and $R_y$ are the rotation gates, $\oplus$ denotes the C-NOT gates and $\dagger$ represents the reverse operations}
    \label{fig:circuit}
\end{figure}

To map features into qubits for building kernel circuits, we employed two rotation gates, $R_x$ and $R_y$. Given two classical features $x_1$ and $x_2$, they can be encoded into the quantum state of a qubit using these rotation gates. The $R_x$ gate applies a rotation about the x-axis of the Bloch sphere, mathematically defined as $R_x(\theta) = e^{-i\theta X/2}$, where $X$ is the Pauli-X matrix and $\theta$ is the angle of rotation, which can be set as proportional to $x_1$. Similarly, the $R_y$ gate rotates the qubit around the y-axis and is defined as $R_y(\phi) = e^{-i\phi Y/2}$, with $Y$ being the Pauli-Y matrix and $\phi$ proportional to $x_2$. By applying $R_x(x_1)$ followed by $R_y(x_2)$ to each qubit, the two features $x_1$ and $x_2$ are mapped to the qubit's quantum state, creating a parameterized quantum state that can be used in quantum kernels. This process encodes classical information into quantum circuits, allowing for quantum-enhanced classical operations. Fig~\ref{fig:circuit} shows the quantum circuit that maps the feature into qubits and creates the quantum kernel. The encoded data goes through a series of CNOT (Controlled-NOT) gates. These gates create entanglement between pairs of qubits, which allows the quantum circuit to capture correlations between the data features. Entanglement is a key resource in quantum computation, as it enables the circuit to represent more complex patterns and relationships in the data. After the entangling operations, the circuit applies inverse operations to effectively complete the quantum feature mapping. Following this, the qubits are measured on a computational basis, yielding classical data as the output. This output is used to calculate the similarity between two data points in the quantum feature space, which forms the quantum kernel.

\subsection{Adversarial Attack with Threat Model}
\begin{figure}
    \begin{subfigure}[h]{0.48\linewidth}
    \includegraphics[width=\linewidth]{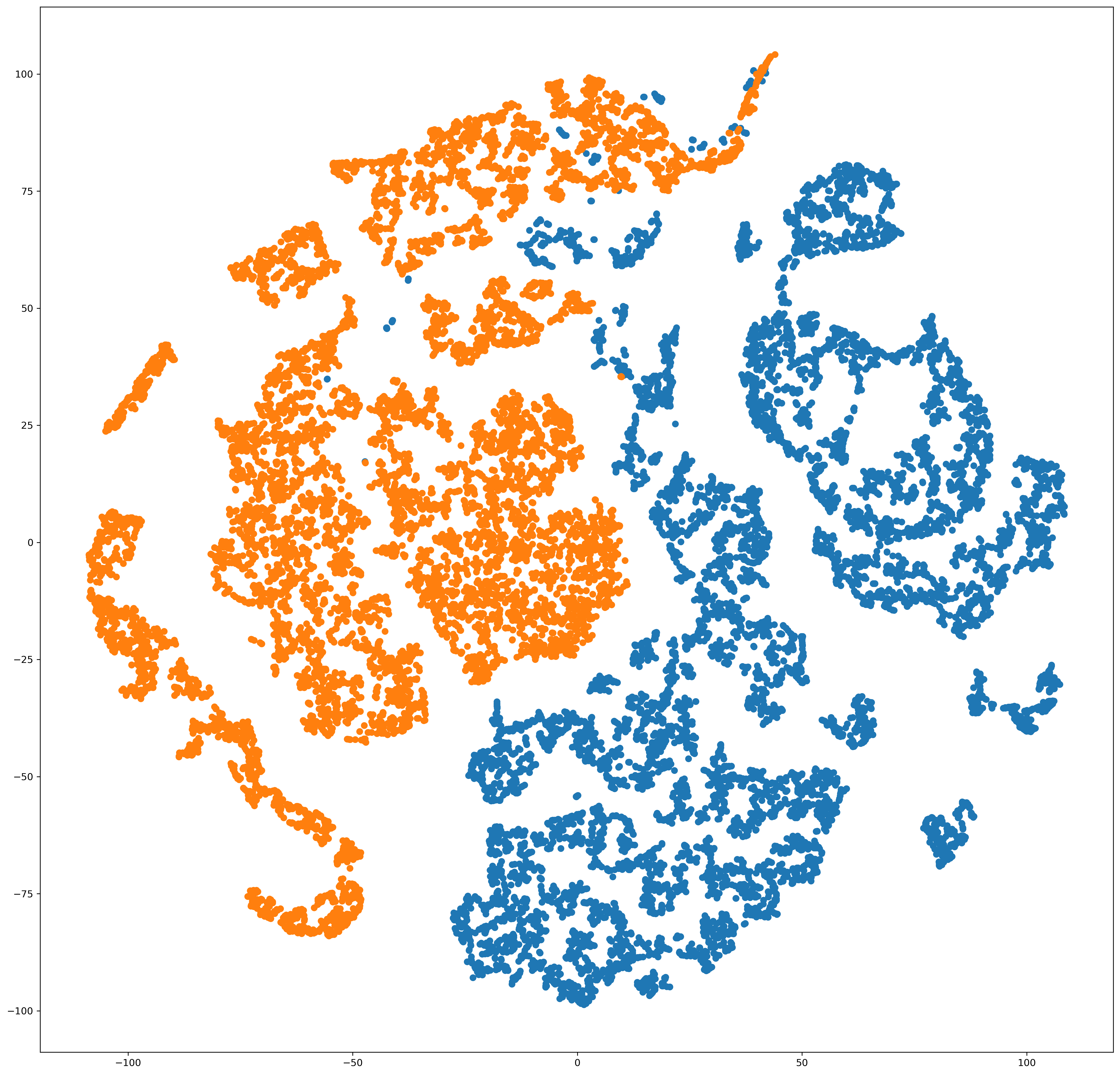}

    \caption{Clean Dataset before the Attack}
    \label{fig:eph_clean}
    \end{subfigure}
\hfill
    \begin{subfigure}[h]{0.48\linewidth}
    \includegraphics[width=\linewidth]{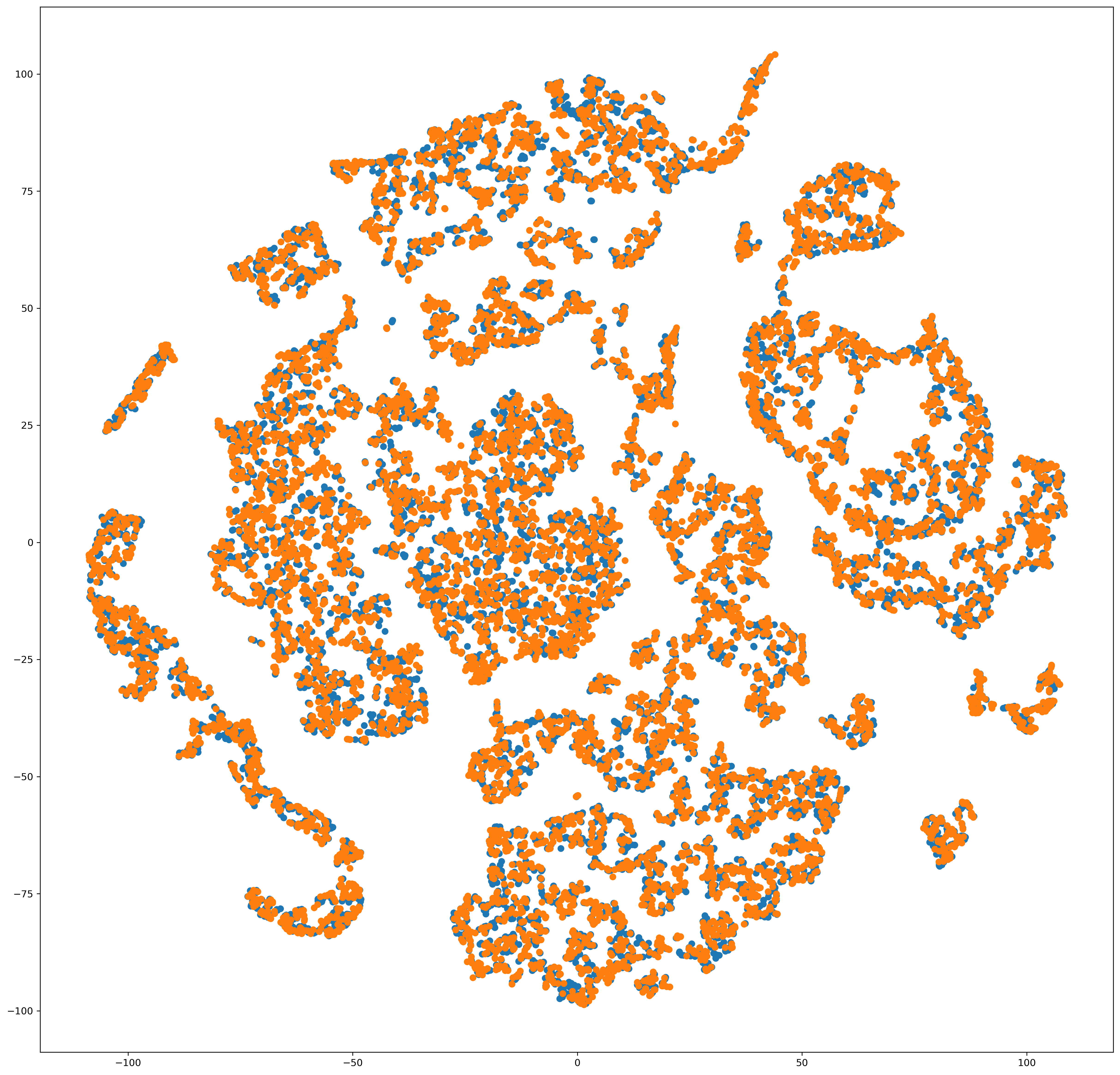}

    \caption{Adversarial Dataset after the Attack}
    \label{fig:eph_attack}
    \end{subfigure}%
\caption{2D Projection of EPHNOGRAM Dataset before and after Data Poisoning Attack where \textit{blue} and \textit{orange} Represent Two Different Classes}
\label{fig:eph_vis}
\end{figure}

We implemented white box attacks on all three datasets and monitored the change in performance for a base model. We see a significant drop in performance in the base model in terms of accuracy as seen in Table~\ref{table:attack_performance}. The metrics are calculated from the SVM dot product kernel averaging 5-fold of the validation set. As a visual example, Fig~\ref{fig:eph_vis} shows the 2D projection of the EPHNOGRAM dataset before and after the label-flipping attack. The blue and orange markers represent two distinct classes fairly separable in Fig~\ref{fig:eph_clean} but after the attack, they are no longer distinguishable as seen in Fig~\ref{fig:eph_attack}.

\begin{table}[htbp]
\caption{Average Performance of Dot Product Kernel SVM for 5-fold Data before and after Adversarial Attack}
\label{table:attack_performance}
\centering
\resizebox{\columnwidth}{!}{
    \setlength{\tabcolsep}{2pt}
    \renewcommand*{\arraystretch}{1.15}
    \begin{tabular}{ccccc}
    \toprule
     \bf Dataset & \bf Class & \makecell[c]{\bf Accuracy(\%)\\ \bf (Clean Data)} & \bf \makecell[c]{Attack\\ \bf Type}  & \bf \makecell[c]{\bf Accuracy($\%\downarrow$) \\ \bf (Adv Data)}\\
        \midrule
        \multirow{2}{*}{RESTING ECG} & \multirow{2}{*}{One} & \multirow{2}{*}{85.009} & Label Flipping & 63.205($\downarrow$21.804) \\
        \cmidrule{4-5}
        & & & Adv Perturbation & 83.745($\downarrow$1.264) \\
        \midrule
        
        \multirow{3}{*}{EPHNOGRAM} & \multirow{3}{*}{Two} & \multirow{3}{*}{99.975} & Label Flipping & 49.556($\downarrow$50.419) \\
        \cmidrule{4-5}
        & & & Targeted Poisoning & 30.525($\downarrow$69.45) \\
        \cmidrule{4-5}
        & & & Adv Perturbation & 62.333($\downarrow$37.642) \\
        \midrule

        \multirow{3}{*}{Stress} & \multirow{3}{*}{Three} & \multirow{3}{*}{80.752} & Label Flipping & 43.878($\downarrow$36.874) \\
        \cmidrule{4-5}
        & & & Targeted Poisoning & 47.788($\downarrow$32.964) \\
        \cmidrule{4-5}
        & & & Adv Perturbation & 75.764($\downarrow$4.987) \\
        \bottomrule
    \end{tabular}
    }
\end{table}

\subsection{Tampering Detection with One-Class SVM}
One-Class SVM works by finding the optimal hyperplane that best separates the samples from potential outliers. Using a kernel can elevate the data to higher dimensional feature space making it easier to draw decision boundaries. The purpose of training phase is to find the optimal hyperplane with the biggest margin that divides the data points of different classes. In kernel-based SVMs, the optimization problem is addressed under the feature space generated by the kernel function of choice. Additionally, the data points closest to the decision boundary's hyperplane are known as support vectors, which SVM learns during training and is used to define the decision boundary. The driving function for a One-Class SVM can be represented in Equation~\eqref{obj_func}.

\begin{equation}
    f(x) = sign\left( \sum_{i=1}^N\sum_{j=1}^N  \alpha_{ij} \times k(x_i, x_j) + b - \rho \right)
    \label{obj_func}
\end{equation}

Here, $\{x_i\}_{i=1}^N$ is training data from one class, $k(x_i, x_j)$ is the dot product kernel. $\alpha_{ij}$ and $b$ are the constants used as a multiplier and bias term and $\rho$ is the threshold. The optimization is done to maximize the margin between the decision boundary represented as Equation~\eqref{opt} for a given constraint mentioned in Equation~\eqref{given_cons}.

\begin{equation}
    \underset{\alpha, \rho}{maximize} = \frac{1}{2} \times ||\omega||^2 - \rho + \frac{1}{\nu \times N} \sum_{i=1}^N{\zeta_i}
    \label{opt}
\end{equation}

\begin{equation}
    when, \omega^T \times \phi(x_i) + b \geq \rho - \zeta_i
    \label{given_cons}
\end{equation}

Here, $\omega$ is the vector normal to the decision hyperplane, $\phi(x_i)$ is the basis function, $\zeta$ is the relaxation value, and $\nu$ controls the trade-off between maximizing the margin and controlling the fraction of outliers. This optimization problem is solved iteratively with gradient descent or sequential minimal optimization. In our work, we used \textit{scikit-learn} module's \textit{svm} class to implement this.

\section{Result Analysis and Discussion}
\label{result}

After the attack, we evaluated the performance of our hybrid model to detect the tampered data points with 5-fold cross-validation and reported the average of the folds. For datasets with a single label (RESTINGECG), the detection accuracy measures the ratio of altered labels or anomalies and the total number of samples. On the other hand, for the datasets with binary (EPHNOGRAM) and multiclass (Stress) labels, the accuracy was calculated class-wise and averaged on 5-folds. Table~\ref{table:performance} shows the performance of detection after each type of attack for the three datasets. It also shows the metrics in comparison to the classical version of the framework.

\begin{table}[htbp]
\caption{Detection Accuracy for Test Set of Proposed Framework after 5-fold Cross validation}
\label{table:performance}
\centering
\resizebox{\columnwidth}{!}{
    \setlength{\tabcolsep}{3pt}
    \renewcommand*{\arraystretch}{1}
    \begin{tabular}{cccccc}
    \toprule
     \bf Method & \bf Dataset & \bf Class & \bf Features & \bf Attack Type & \bf Detection (\%)\\
        \midrule
        \multirow{8}{*}{\rotatebox[origin=c]{90}{Classical}} & \multirow{2}{*}{RESTING ECG} & \multirow{2}{*}{One} & \multirow{2}{*}{12}  & Label Flipping & 67.67\\
        \cmidrule{5-6}
        & & & & Adv Perturbation & 60.52\\
        \cmidrule{2-6}
        & \multirow{3}{*}{EPHNOGRAM} & \multirow{3}{*}{Two} & \multirow{3}{*}{12} & Label Flipping & 75.01\\
        \cmidrule{5-6}
        & & & & Targeted Poisoning & 50.36\\
        \cmidrule{5-6}
        & & & & Adv Perturbation & 68.99\\
        \cmidrule{2-6}
        & \multirow{3}{*}{Stress} & \multirow{3}{*}{Three} & \multirow{3}{*}{12} & Label Flipping & 80.75\\
        \cmidrule{5-6}
        & & & & Targeted Poisoning & 65.23\\
        \cmidrule{5-6}
        & & & & Adv Perturbation & 58.95\\
        \midrule
        \multirow{9}{*}{\rotatebox[origin=c]{90}{Quantum}} & \multirow{2}{*}{RESTING ECG} & \multirow{2}{*}{One} & \multirow{2}{*}{12}  & Label Flipping & 75.03\\
        \cmidrule{5-6}
        & & & & Adv Perturbation & 55.16\\
        \cmidrule{2-6}
        & \multirow{3}{*}{EPHNOGRAM} & \multirow{3}{*}{Two} & \multirow{3}{*}{12} & Label Flipping & 75.23\\
        \cmidrule{5-6}
        & & & & Targeted Poisoning & 66.66\\
        \cmidrule{5-6}
        & & & & Adv Perturbation & 60.09\\
        \cmidrule{2-6}
        & \multirow{3}{*}{Stress} & \multirow{3}{*}{Three} & \multirow{3}{*}{12} & Label Flipping & 95.11\\
        \cmidrule{5-6}
        & & & & Targeted Poisoning & 49.96\\
        \cmidrule{5-6}
        & & & & Adv Perturbation & 45.99\\
        \bottomrule
    \end{tabular}
    }
\end{table}

From the table, it is seen that both classical and quantum SVM performed better against label-flipping attacks. In terms of the datasets, the hybrid model performed about 12\% better for RESTINGECG single-label data against label-flipping. For EPHNOGRAM with two labels, detection of label-flipping attacks is similar for both but the hybrid model performed better against targeted poisoning attacks. Finally, in the Stress dataset with three labels, the proposed hybrid model has 15\% more accuracy against label-flipping attacks. Their performance on Adversarial perturbation is not satisfactory at all. FGSM-based adversarial perturbations are inherently harder to detect as the subtle perturbation also keeps the statistical properties intact. Thus both classical and hybrid models perform poorly against them.

\section{Conclusion and Future Work}
\label{Conclusion}
In this work, we explored the performance of hybrid QML for tampering detection. Our experiment on 3 types of datasets makes our findings more viable for real-world deployment. Across all datasets, quantum models consistently achieved higher detection rates for label-flipping attacks compared to classical models with a maximum of 15\% increased accuracy. This suggests that quantum algorithms are better suited for identifying data poisoning attacks. While both models struggle with adversarial perturbation attacks, quantum models demonstrate a slight improvement over classical models in terms of detection accuracy. This suggests that quantum algorithms may offer some advantages in handling more sophisticated attacks. This highlights the need for further research to improve model resilience against these types of attacks in the context of physiological data security. One of the research directions would be to explore more robust machine learning models like Quantum Neural Networks (QNN). Again, different kinds of kernel circuits may provide insight into better kernel and fidelity calculation resulting in better detection accuracy. 

\bibliographystyle{style}
\bibliography{reference}\balance

\begin{thebibliography}{10}
\providecommand{\url}[1]{#1}
\csname url@samestyle\endcsname
\providecommand{\newblock}{\relax}
\providecommand{\bibinfo}[2]{#2}
\providecommand{\BIBentrySTDinterwordspacing}{\spaceskip=0pt\relax}
\providecommand{\BIBentryALTinterwordstretchfactor}{4}
\providecommand{\BIBentryALTinterwordspacing}{\spaceskip=\fontdimen2\font plus
\BIBentryALTinterwordstretchfactor\fontdimen3\font minus
  \fontdimen4\font\relax}
\providecommand{\BIBforeignlanguage}[2]{{%
\expandafter\ifx\csname l@#1\endcsname\relax
\typeout{** WARNING: IEEEtran.bst: No hyphenation pattern has been}%
\typeout{** loaded for the language `#1'. Using the pattern for}%
\typeout{** the default language instead.}%
\else
\language=\csname l@#1\endcsname
\fi
#2}}
\providecommand{\BIBdecl}{\relax}
\BIBdecl

\bibitem{malik2024}
J.~Malik, R.~Muthalagu, and P.~M. Pawar, ``A systematic review of adversarial
  machine learning attacks, defensive controls, and technologies,'' \emph{IEEE
  Access}, vol.~12, pp. 99\,382--99\,421, 2024.

\bibitem{li2010}
M.~Li, W.~Lou, and K.~Ren, ``Data security and privacy in wireless body area
  networks,'' \emph{IEEE Wireless Communications}, vol.~17, no.~1, pp. 51--58,
  2010.

\bibitem{dimitriou2008}
T.~Dimitriou and K.~Ioannis, ``Security issues in biomedical wireless sensor
  networks,'' in \emph{2008 First International Symposium on Applied Sciences
  on Biomedical and Communication Technologies}, 2008, pp. 1--5.

\bibitem{pathak2021}
A.~K. Pathak, S.~Saguna, K.~Mitra, and C.~Åhlund, ``Anomaly detection using
  machine learning to discover sensor tampering in iot systems,'' in \emph{ICC
  2021 - IEEE International Conference on Communications}, 2021, pp. 1--6.

\bibitem{pacheo2018}
J.~Pacheco and S.~Hariri, ``Anomaly behavior analysis for iot sensors,''
  \emph{Transactions on Emerging Telecommunications Technologies}, vol.~29,
  no.~4, p. e3188, 2018, e3188 ETT-16-0338.R1.

\bibitem{montalvo2023}
P.~D. Rosero-Montalvo, Z.~István, P.~Tözün, and W.~Hernandez, ``Hybrid
  anomaly detection model on trusted iot devices,'' \emph{IEEE Internet of
  Things Journal}, vol.~10, no.~12, pp. 10\,959--10\,969, 2023.

\bibitem{sun2022}
G.~Sun, Y.~Cong, J.~Dong, Q.~Wang, L.~Lyu, and J.~Liu, ``Data poisoning attacks
  on federated machine learning,'' \emph{IEEE Internet of Things Journal},
  vol.~9, no.~13, pp. 11\,365--11\,375, 2022.

\bibitem{yerlikaya2022}
F.~A. Yerlikaya and Şerif Bahtiyar, ``Data poisoning attacks against machine
  learning algorithms,'' \emph{Expert Systems with Applications}, vol. 208, p.
  118101, 2022.

\bibitem{qml}
M.~Schuld and N.~Killoran, ``Quantum machine learning in feature hilbert
  spaces,'' \emph{Phys. Rev. Lett.}, vol. 122, p. 040504, Feb 2019.

\bibitem{Havlicek2019}
V.~Havl{\'i}{\v{c}}ek, A.~D. C{\'o}rcoles, K.~Temme, A.~W. Harrow, A.~Kandala,
  J.~M. Chow, and J.~M. Gambetta, ``Supervised learning with quantum-enhanced
  feature spaces,'' \emph{Nature}, vol. 567, no. 7747, pp. 209--212, Mar 2019.

\bibitem{cultice2024}
T.~Cultice, M.~S.~H. Onim, A.~Giani, and H.~Thapliyal, ``Anomaly detection for
  real-world cyber-physical security using quantum hybrid support vector
  machines,'' in \emph{2024 IEEE Computer Society Annual Symposium on VLSI
  (ISVLSI)}.\hskip 1em plus 0.5em minus 0.4em\relax IEEE, 2024, pp. 619--624.

\bibitem{xu2012}
D.~Xu, M.~Tu, M.~Sanford, L.~Thomas, D.~Woodraska, and W.~Xu, ``Automated
  security test generation with formal threat models,'' \emph{IEEE Transactions
  on Dependable and Secure Computing}, vol.~9, no.~4, pp. 526--540, 2012.

\bibitem{moosavi2017universal}
S.-M. Moosavi-Dezfooli, A.~Fawzi, O.~Fawzi, and P.~Frossard, ``Universal
  adversarial perturbations,'' in \emph{Proceedings of the IEEE conference on
  computer vision and pattern recognition}, 2017, pp. 1765--1773.

\bibitem{restingecg}
L.~T. Trujillo, C.~T. Stanfield, and R.~D. Vela, ``The effect of
  electroencephalogram (eeg) reference choice on information-theoretic measures
  of the complexity and integration of eeg signals,'' \emph{Frontiers in
  Neuroscience}, vol.~11, 2017.

\bibitem{ephnogram}
A.~Kazemnejad, P.~Gordany, and R.~Sameni, ``Ephnogram: A simultaneous
  electrocardiogram and phonocardiogram database,'' 2021.

\bibitem{PhysioBank}
A.~L. Goldberger, L.~A.~N. Amaral, L.~Glass, J.~M. Hausdorff, P.~C. Ivanov,
  R.~G. Mark, J.~E. Mietus, G.~B. Moody, C.-K. Peng, and H.~E. Stanley,
  ``Physiobank, physiotoolkit, and physionet: Components of a new research
  resource for complex physiologic signals,'' \emph{Circulation}, vol. 101,
  no.~23, Jun. 2000.

\bibitem{nath2022}
R.~K. Nath, H.~Thapliyal, and A.~Caban-Holt, ``Machine learning based stress
  monitoring in older adults using wearable sensors and cortisol as stress
  biomarker,'' \emph{Journal of Signal Processing Systems}, pp. 1--13, 2022.

\bibitem{onim2024utilizing}
M.~S.~H. Onim, H.~Thapliyal, and E.~K. Rhodus, ``Utilizing machine learning for
  context-aware digital biomarker of stress in older adults,''
  \emph{Information}, vol.~15, no.~5, p. 274, 2024.

\bibitem{onim2024predict}
M.~S.~H. Onim and H.~Thapliyal, ``Predicting stress in older adults with rnn
  and lstm from time series sensor data and cortisol,'' in \emph{2024 IEEE
  Computer Society Annual Symposium on VLSI (ISVLSI)}, 2024, pp. 300--306.

\bibitem{pca}
F.~L. Gewers, G.~R. Ferreira, H.~F.~D. Arruda, F.~N. Silva, C.~H. Comin, D.~R.
  Amancio, and L.~D.~F. Costa, ``Principal component analysis: A natural
  approach to data exploration,'' vol.~54, no.~4, May 2021.

\end{thebibliography}

\end{document}